\documentclass[twocolumn]{emulateapj}

\begin{document}

\title{Outgassing Behavior of C/2012 S1 (ISON) from 2011 September to 2013 June}

\author{
Karen J.\ Meech\altaffilmark{1,2},
Bin Yang\altaffilmark{1,2},
Jan Kleyna\altaffilmark{1,2},
Megan Ansdell\altaffilmark{2},
Hsin-Fang Chiang\altaffilmark{1,2},
Olivier Hainaut\altaffilmark{3},
Jean-Baptiste Vincent\altaffilmark{4},
Hermann Boehnhardt\altaffilmark{4},
Alan Fitzsimmons\altaffilmark{5},
Travis Rector\altaffilmark{6},
Timm Riesen\altaffilmark{1,2},
Jacqueline V.\ Keane\altaffilmark{1,2},
Bo Reipurth\altaffilmark{1,2},
Henry H.\ Hsieh\altaffilmark{2,a},
Peter Michaud\altaffilmark{7},
Giannantonio Milani\altaffilmark{8,b},
Erik Bryssinck\altaffilmark{9,b},
Rolando Ligustri\altaffilmark{10,b},
Roberto Trabatti\altaffilmark{11,b},
Gian-Paolo Tozzi\altaffilmark{12},
Stefano Mottola\altaffilmark{13},
Ekkehard Kuehrt\altaffilmark{13},
Bhuwan Bhatt\altaffilmark{14},
Devendra Sahu\altaffilmark{14},
Carey Lisse\altaffilmark{15},
Larry Denneau\altaffilmark{2},
Robert Jedicke\altaffilmark{2},
Eugene Magnier\altaffilmark{2},
Richard Wainscoat\altaffilmark{2}.
}

\altaffiltext{1}{NASA Astrobiology Institute}
\altaffiltext{2}{Institute for Astronomy, University of Hawaii,
2680 Woodlawn Drive, Honolulu, HI 96822, USA}
\altaffiltext{3}{European Southern Observatory}
\altaffiltext{4}{Max-Planck-Institut fur Sonnensystemforschung, Max-Planck-Strasse 2, 37191, Katlenburg-Lindau, Germany}
\altaffiltext{5}{Queens Univ. Belfast, Belfast BT7 1NN, Northern Ireland}
\altaffiltext{6}{Dept. of Physics and Astronomy, University of Alaska Anchorage, 3211 Providence Dr., Anchorage, AK 99508}
\altaffiltext{7}{Gemini Observatory, Northern Operations Center, 670 N.\ AÔohoku Place, Hilo, HI 96720, USA}
\altaffiltext{8}{Associazione Astronomica Euganea, via Tommaseo, 35131 Padova, Italy}
\altaffiltext{9}{BRIXIIS Observatory, Eyckensbeekstraat, 9150 Kruibeke, Belgium}
\altaffiltext{10}{Talmassons Observatory (C.A.S.T.), via Cadorna, 33030 Talmassons, Italy}
\altaffiltext{11}{Stazione Astronomica Descartes, via Lambrinia 4, 2013 Chignolo Po', Italy}
\altaffiltext{12}{INAF-Osservatorio Astrofisico di Arcetri, Largo E. Fermi 5, I-40125 Firenze, Italy}
\altaffiltext{13}{DLR--German Aerospace Center, Institute of Planetary
Research, Rutherfordstr. 2, D-12489 Berlin, Germany}
\altaffiltext{14}{Indian Inst. Astrophys., II Block, Koramangala, Bangalor 560 034, India}
\altaffiltext{15}{JHU-APL, 11100 Johns Hopkins Road, Laurel MD 20723}
\altaffiltext{a}{Hubble Fellow}
\altaffiltext{b}{CARA Project, Astrofili Italiani, IASF INAF via Fosso del Cavaliere 100, 00133 Roma, Italy}

\email{meech@ifa.hawaii.edu}

\slugcomment{Submitted, 2013-08-25; Revised, 2013-09-07; Accepted, 2013-09-09}

\begin{abstract}

We report photometric observations for comet C/2012 S1 (ISON)
obtained during the time period immediately after discovery ($r$=6.28
AU) until it moved into solar conjunction in mid-2013 June
using the UH2.2m, and Gemini North 8-m telescopes on Mauna Kea, the
Lowell 1.8m in Flagstaff, the Calar Alto 1.2m telescope in Spain,
the VYSOS-5 telescopes on Mauna Loa Hawaii and data from the CARA
network. Additional pre-discovery data from the Pan STARRS1 survey
extends the light curve back to 2011 September 30 ($r$=9.4 AU). The
images showed a similar tail morphology due to small
micron sized particles throughout 2013.  Observations at sub-mm
wavelengths using the JCMT on 15 nights between 2013 March 9 ($r$=4.52
AU) and June 16 ($r$=3.35 AU) were used to search for CO and
HCN rotation lines. No gas was detected, with upper limits for CO
ranging between 3.5-4.5$\times$10$^{27}$ molec s$^{-1}$. Combined with 
published water production rate estimates we have generated ice sublimation
models consistent with the photometric light curve. The inbound
light curve is likely controlled by sublimation of CO$_2$.  At these
distances water is not a strong contributor to the outgassing. We
also infer that there was a long slow outburst of activity beginning
in late 2011 peaking in mid-2013 January (r$\sim$5 AU) at which point
the activity decreased again through 2013 June.  We suggest
that this outburst was driven by CO injecting large water ice
grains into the coma. Observations as the comet came out of solar
conjunction seem to confirm our models.

\end{abstract}


\newpage

\section{INTRODUCTION}

On 2012 September 21 a new sungrazing comet was discovered using
the 0.4-meter International Scientific Optical Network telescope
in Russia \citep{nevski12}. The comet was designated C/2012 S1
(ISON) (hereafter comet ISON) and was bright and active at 6.3 AU
pre-perihelion.  The current estimate of its orbital eccentricity
is 1.000004, thus it is possibly making its first passage through
the inner solar system from the Oort cloud. Perihelion is on 2013
November 28 at a distance of 0.0125 AU (2.7 solar radii), and some
predictions suggest it could become exceedingly bright.  About a
dozen comets in the past $\sim$270 years have been spectacularly
bright (mag$<$-5), and the hope that comet ISON could be one of
these has generated intense scientific interest.  However, it is
difficult to predict the comet's behavior while still far from the Sun.
Comet ISON was well placed for observation until moving into solar
conjuction in 2013 June, and it emerged again in the dawn skies in
late 2013 August, near r=2.4 AU. In this letter, we report observations
of the comet at optical and submillimeter wavelengths from 2011 September
through 2013 June.  Based on these data and the gas production
rates from the literature, we used an ice sublimation model to look
at activity scenarios for when the comet emerged from solar
conjunction.

\section{OBSERVATIONS \& DATA REDUCTION\label{observations}}

We initiated both a pre-perihelion imaging campaign and a sub-mm
observing campaign to constrain volatile production rates (see
Table~\ref{obslog}).  Imaging data were taken on both photometric
nights and nights with some cirrus.  Calibrations on photometric
nights were accomplished with measurements of \citet{landolt92}
standard stars.  Fields on non-photometric nights and for Pan-STARRS1
(PS1) were calibrated against the Sloan Digital Sky Survey (SDSS;
\citet{york00}) or the PS1 2-pi survey \citep{magnier13}. Conversion
to Kron-Cousins R-band used the transformation equations derived
by R.  Lupton\footnote{http://www.sdss.org/} and \citet{tonry12}
assuming an observed color of $V-R$=0.4 \citep{lisse13}.

\subsection{Pan STARRS1}

Comet ISON was detected in images obtained with PS1
and the Gigapixel Camera 1 (0.256$''$ pixels) between 2011 September
and 2013 January during regular survey operations.
Exposures were made in the survey $grizw_{P1}$ filters. Moving
objects are normally automatically detected and measured via
difference imaging \citep{denneau13}.  Before 2012 January the comet
was moving too slowly and/or was too faint for this to be successful;
these detections were made by manual inspection of the data
post-discovery.  In all PS1 prediscovery data the comet has a profile
no wider than the point spread function (PSF) of field stars,
although we infer it was likely to be active at this time (see
\S\ref{smodel}).  The magnitudes were measured via DAOPHOT
PSF-photometry relative to field stars of known magnitudes
\citep{schlafly12}.  The 2012 January 28 detections were reported
to the Minor Planet Center (MPC) within 24 hours. However as its
motion was roughly parallel to the ecliptic, and the PSF was measured
to be stellar, it was not reported as an object of interest.  From
2012 September onwards the comet possessed a visible coma, and
magnitudes were measured in the $PS1$ photometric system within a
5$''$ radius circular aperture \citep{schlafly12, magnier13} when
not contaminated by field stars.

\subsection{UH2.2m, HCT2m, Lowell Observatory 1.8m, and Calar Alto 1.2m}

We had several observing runs in 2012-2013 where we obtained data on
comet ISON.  The UH2.2m telescope on Mauna Kea was used with the
Tek 2K CCD camera and Kron-Cousins filter.  Data were obtained using
the Himalayan Chandra telescope with the Optical 2K$\times$4K imager
(E2V chips with image scale of 0.17$''$ pix$^{-1}$) through Bessell
R-band filters.  Observations at the Lowell observatory
were taken on the Perkins 1.8m telescope with the PRISM reimaging
camera (SITE 2K$\times$2K CCD) through the Bessell R band. Data
were obtained under variable conditions during several nights using
the Calar Alto 1.2m with the e2v CCD binned 2$\times$2.

\subsection{Gemini North 8-m}

\begin{figure}
\includegraphics[width=18cm]{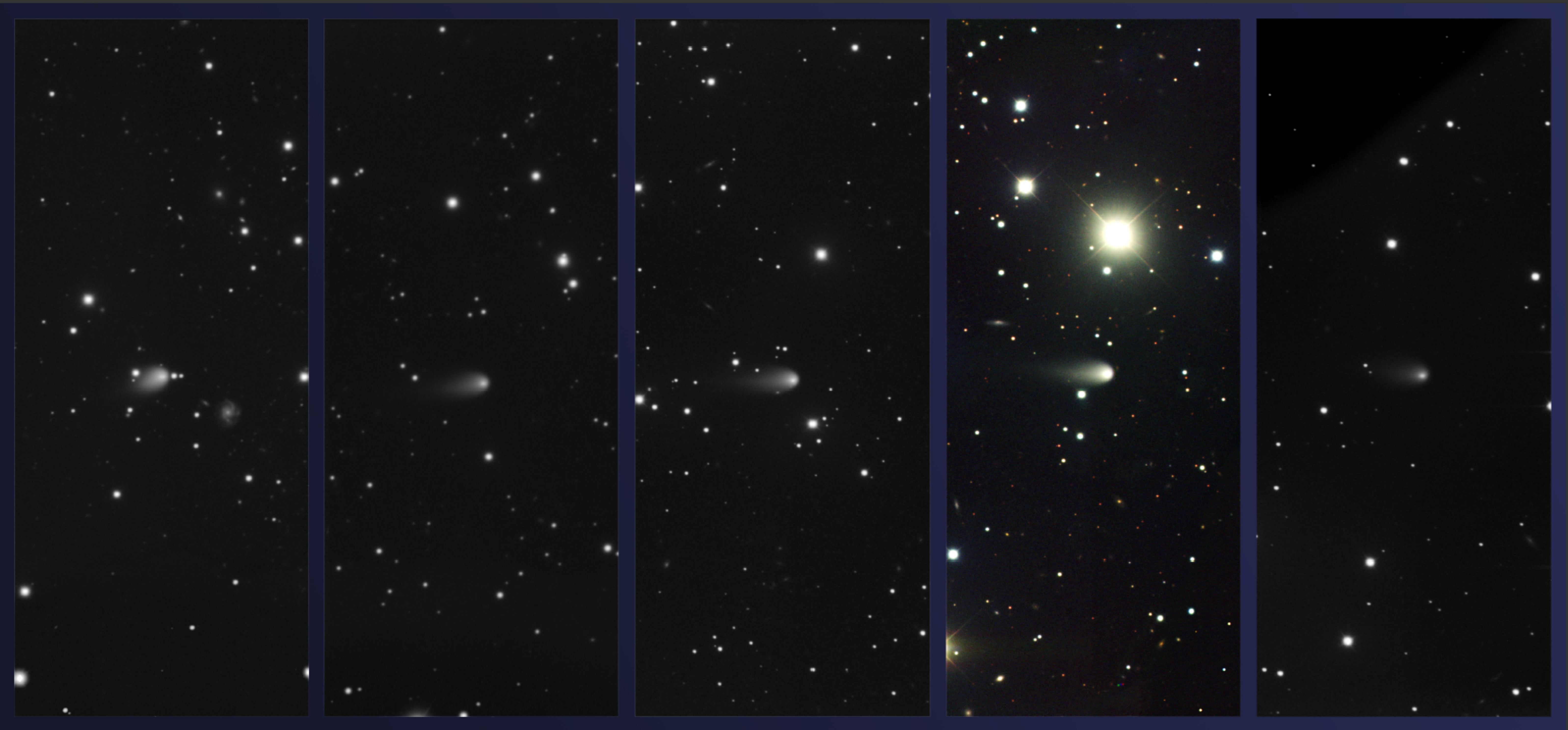}
\caption{\small 
Images of comet ISON obtained using the Gemini 8m with GMOS on 2013 
February 4, March 4, April 3, May 4 and 30 through an $r$ filter.  The 
May 4 image is a color composite image made using the $g$, $r$ and $i$ images.
East is left, N is up and the FOV is $\sim$2.5$'$ wide per panel
(corresponding to a projected distance of 4.25-4.70$\times$10$^5$
km at the distance of the comet).
}
\label{fig-Gemini}
\end{figure}

Director's discretionary time was awarded to image the comet monthly
from 2013 February through June (see Fig.~\ref{fig-Gemini}).  Data
were obtained using the GMOS-N multi-object spectrograph in imaging
mode with the e2V CCD binned 2$\times$2 resulting in a plate scale
of 0.145$''$ pix$^{-1}$.  Baseline calibrations were used in the
reduction. Most of the images were obtained through the Sloan
r$'$-filter \citep{fukugita96}.  Some non-photometric nights were
calibrated using the USNO-A2 catalog, resulting in a photometric
accuracy of 5-10\%.

\subsection{JCMT}

The radio observations were performed using the 15-m JCMT telescope
on 15 days between 2013 March 9 through 2013 June 16.  Long integration
observations were performed on March 15 and 30, April 1, 27-29 and
June 14-16.  At other times, hourly snapshot observations were
obtained. We used HARP and the R$\times$A3 heterodyne receivers in
beam-switch ({\it i.e.} secondary chopping) mode. The planets Mars
and Uranus were frequently observed to monitor the main beam
efficiency. The ACSIS spectrometer was used, which provides a total
bandwidth of 250 MHz and a spectral channel spacing of 30.5 kHz.
The data were reduced using a combination of the Star-link software
and IDL scripts.

Searches for the J=3-2 and J=2-1 rotational transitions of CO gas
and the J=4-3 and J=3-2 transitions of HCN gas returned negative
results.  HCN is an indicator of sublimated gas, but not expected
to play a major role in controlling the brightness of the comet,
so we focus only on the analysis of CO.  To estimate production
rates of CO, we measured the root-mean-square value of the main
beam brightness temperature fluctuations within $\pm$10 km s$^{-1}$
of zero velocity (see Table~\ref{obslog}).  Given that CO lines are
likely to be narrow \citep{senay94,biver02}, the 3-$\sigma$ upper
limits to the line area were derived within a 1.2 km s$^{-1}$ or
1.5 km s$^{-1}$ (for June only) band. We assume that gas molecules
escape from the surface at a constant velocity and follow a Haser
density distribution. We adopted an average expansion velocity of
1.12$\cdot$r$^{-0.41}$ km s$^{-1}$ and the kinetic temperature was
estimated using an empirical formula of 116$\cdot$r$^{-1.24}$ K,
where $r$ is the heliocentric distance \citep{biver97}. The derived
production rate upper limits, for the CO J(2-1) and J(3-2) transitions,
are listed in Table~\ref{obslog}.

\subsection{The CARA Project}

CARA is a consortium of amateur astronomers who have developed a
standardized approach to observing comets.  Photometry through a
Cousins R-filter was obtained on 46 dates (Table~\ref{obslog2})
beginning shortly after discovery in 2012 Septemper through 2013 May
2 with most of the observations coming from 0.4m telescopes at
the BRIXIIS Observatory in Belgium, the Talmassons Observatory and
Stazione Astronomica Descartes in Italy.  The photometry was
calibrated using the APASS catalog\footnote{http://www.aavso.org/apass}.

\subsection{VYSOS Telescope\label{vysos}}

We used the 5.3 inch Variable Young Stellar Objects Survey (VYSOS)
program\footnote{http://www.ifa.hawaii.edu/\~{}reipurth/VYSOS/}
robotic refractor at the Mauna Loa Observatory in Hawai'i, with an
Apogee Alta U16M CCD (field of view 2.9$\times$2.9 degrees with a
plate scale of 2.53$''$ pixel$^{-1}$) to image the comet.  Images
were taken nearly nightly from 2013 April to mid-June
(Table~\ref{obslog2}).  On most nights, at least three exposures
of 100 sec each were taken in a Sloan r$'$-band filter.

We used Sextractor\footnote{http://www.astromatic.net} to extract
detections from our 165 VYSOS images, using an aperture that contained
$>$95\% of the PSF.  We corrected for focal plane irregularities
with an approximate distortion map, permitting us to match stellar
detections between frames in the VYSOS run.  We used the first image
to perform relative photometric calibrations of subsequent images,
and used overlapping stars from frame to frame for calibration--thus
establishing a relative zero-point calibration spanning the entire
run has a nominal uncertainty of 0.027 mag.  For some VYSOS frames,
the survey overlapped with SDSS, permitting us to calibrate one
selected frame to 0.014 mag using 39 SDSS stars, establishing an
absolute magnitude scale for the entire run.  ISON's ephemeris was
used to search for it in the Sextractor catalogs and then measure
it, finding an object within 4$''$ ($<$2 pixels) of its predicted
location in 129 frames (the others were excluded because of chip
gaps and field star proximity).


\section{ANALYSIS \& RESULTS\label{results}}

\subsection{Finson-Probstein Dust Modeling\label{fpmodel}}

Finson-Probstein modeling \citep{finson68} was used to analyze the
synchrone-syndyne pattern and the optical appearance of the comet's dust
environment (modeling details are described in
\citet{beisser87, vincent10}). Due to projection, as seen from Earth
the syndyne-synchrone pattern converges in the direction of the
dust tail, aligning with the central axis of the optical dust tail.
The travel time of micron-sized dust through the dust tail in the
images is $\sim$20-30 days.  Larger grains of $\sim$100 $\mu$m size
may stay much longer ($\sim$100 days) in the immediate (5$''$ radius)
neighborhood of the nucleus.  The width of the dust tail suggests
a dust expansion speed of $\sim$10 m s$^{-1}$ assuming
micron size dust grains dominate its optical appearance. It is
reasonable to assume that larger (100$\mu$m) grains leave the nucleus
dust acceleration zone at a speed near one to a few m s$^{-1}$.

\subsection{Conceptual Ice Sublimation Model\label{smodel}}

We used a simplified ice sublimation model to investigate the level
of activity versus heliocentric distance.  The model computes the
amount of gas sublimating from an icy surface exposed to solar
heating \citep{meech86,meech04}.  As the ice sublimates, either
from the nucleus surface or near-subsurface, the escaping gas
entrains dust in the flow which escapes into the coma and tail.
The scattered brightness of the comet as measured from Earth has a
contribution from the light scattered from the nucleus and the dust.
Model free parameters include the ice type, nucleus radius, albedo,
emissivity, density, properties of the dust (sizes, density, phase
function), and fractional active area.

%

Our model assumes a nucleus radius of $R_N$$\sim$2 km, consistent
with the size limit inferred from Hubble Space Telescope (HST)
measurements \citep{li13}, an albedo of 0.04 for both the nucleus
and dust, and a linear phase function of 0.04 mag deg$^{-1}$ typical of
other comets.  We assume a nucleus density of 400 kg m$^{-3}$ similar
to that seen for comets 9P/Tempel 1 and 103P/Hartley 2 \citep{thomas13a,
thomas13b}, a grain density of 1000 kg m$^{-3}$, and micron-sized
grains (see $\S$\ref{fpmodel} and \citet{yang13}).  Because of the
many model free parameters, our conclusions are dependent on how
well we can constrain some of the values with observations.  The
shape of the light curve--{\it i.e.} where the curve is steep or
shallow--is determined by the sublimating ice composition.  With
reasonable estimates of nucleus size, albedo, density, and grain
properties, the fractional active surface area is adjusted to produce
the observed volatile production rates. Note, if the HST nucleus
size is much smaller than the upper limit used here, the model will require
an increase in the active fractional area, but otherwise the discussion
below remains unchanged.

\begin{figure}
\includegraphics[width=10cm]{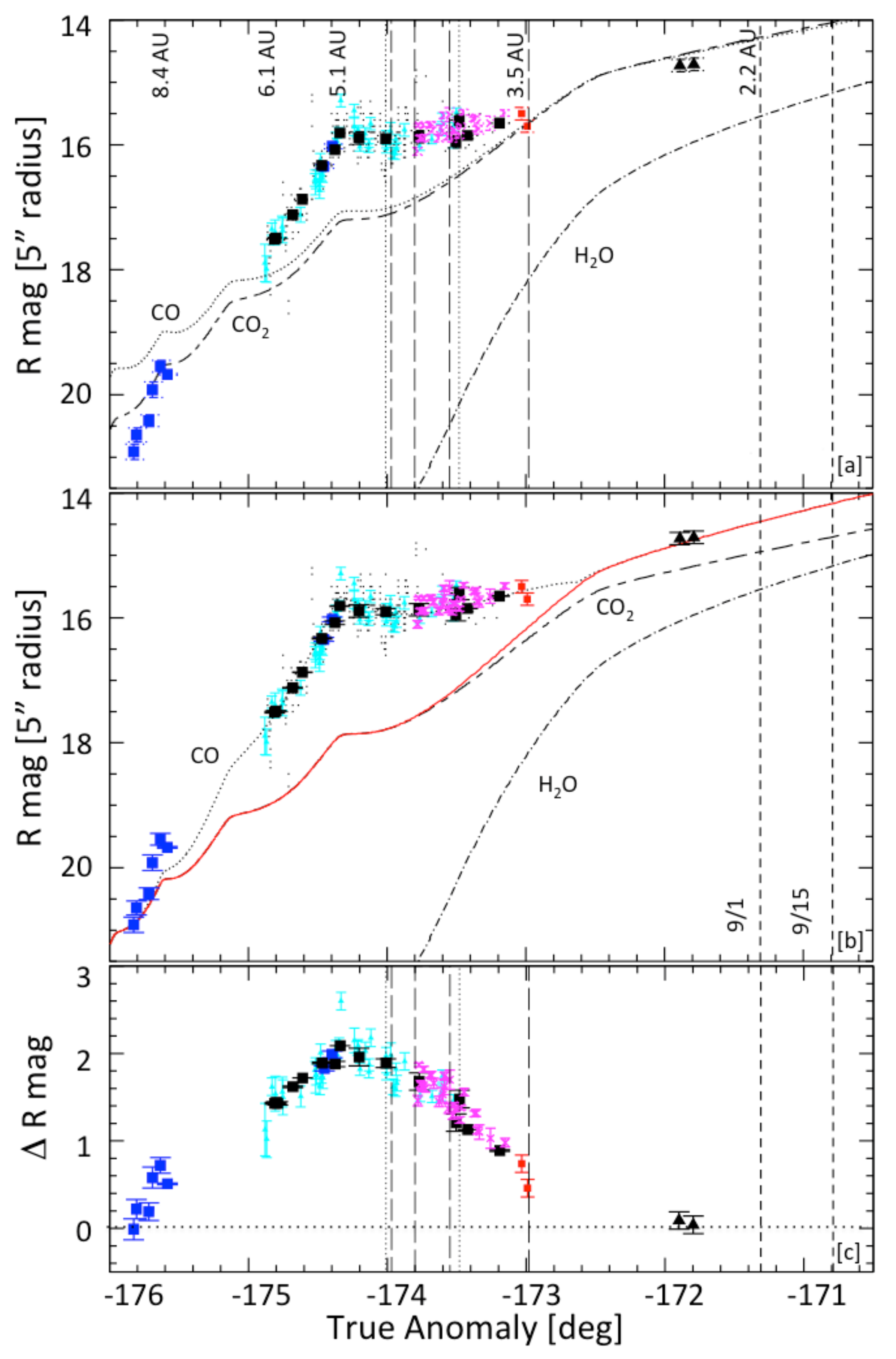}
\caption{\small 
Conceptual ice sublimation model, showing the best fit for 2013 June 13
(TA=-173.0).  The different ice models are for H$_2$O (dot-dash),
CO$_2$ (long-short dash), CO (dotted) and total (solid red line).
The large telescope photometric measurements are shown as large
black squares, the PS1 pre-covery data as blue squares, the CARA
data as cyan triangles, the data from VYSOS as the magenta crosses,
and data as reported in the MPECs as black dots (uncorrected for
aperture size). The optical data obtained at the time of the
{\it Spitzer} observations \citep{lisse13} are shown as small red
squares. The reported measurements by B. Gary are shown as black
triangles. The vertical dotted lines show the dates for which
$Q_{H2O}$ has been published.  The vertical long dashed lines show
the dates where we present JCMT $Q_{CO}$ upper limits and the short
dashed lines indicate when the comet will likely become accessible
to large telescopes as it comes out of solar conjunction.  [a] Best
fit models assuming that all the reported {\it Spitzer} outgassing
is due to only CO or CO$_2$.  [b] Best fit model with baseline
CO$_2$ plus H$_2$O sublimation with an additional slow CO outburst.
[c] Difference between the data and sublimation model, showing the outburst.
}
\label{fig-model}
\end{figure}

Because the comet was active at discovery ($r$=6.3~AU) where it was
too cold for significant water-ice sublimation, there must be another
volatile besides H$_2$O responsible for the outgassing.  The likely
candidates are CO and CO$_2$.  The warm {\it Spitzer} measurements
on June 13 \citep{lisse13} detected an excess brightness at 4.5$\mu$m
due to emission from a neutral gas coma which could either be due
to CO$_2$ or CO (because both have lines in the bandpass).
Unfortunately, there have been no definitive spectral detections
of either molecule reported yet, however the similarity of the
estimated CO$_2$ production rate to measurements of other comets
at large distances \citep{ootsubo12} suggested that CO$_2$ dominated.
We ran two models, assuming the excess seen by {\it Spitzer} was
either all CO or CO$_2$ (see Table~\ref{fluxes}) and the best fit
models are shown in Fig.~\ref{fig-model}a.  While both models can
fit the inferred {\it Spitzer} and water production rates, and match
the scattered light data from the dust in 2013 June, neither model
alone is a good match to the light curve.

Our June CO production upper limits (which agreed with preliminary
estimates from HST, M. A'Hearn, private comm., 2013), also suggested that
the {\it Spitzer} observations were mostly CO$_2$. With this scenario,
however, the only explanation for the light curve between 6-3.5 AU
was a long slow outburst, driven most likely by CO (and supported
by a possible CO detection, N.  Biver, private communication, 2013).  Such
a scenerio is physically plausible, as evidenced by the aperiodic
CO driven outbursts of Comet 29P/Schwassmann Wachmann 1 at similar
distances \citep{cochran82, crovisier95}.

While we were preparing the models, an amateur astronomer, B.
Gary\footnote{brucegary.net/ISON} reported the recovery of the comet
as it came out of solar conjunction on 2013 August 12 and 16 at 7
airmasses.  These magnitudes are also included in Table~\ref{obslog}.
At $r$=2.5 AU, H$_2$O sublimation should be important and we added this
to the model and found that with a fractional active area of 2.5\%
for water sublimation and 0.54\% for CO$_2$, the model fit both the
2013 August data and the early PS1 data with the comet being largely
controlled by CO$_2$ outgassing as shown in Fig.~\ref{fig-model}b.

If we allow that solar heat from the inbound orbit reached a deeper
layer of CO this could trigger additional outgassing starting around
the time of the first PS1 observations. The effect was a period of
increased activity reaching a maximum effective sublimating area
of $\sim$3.4\% at $r$=5.1 AU in 2013 January, and linearly dropping
off and ceasing activity in mid July. This increase is shown as the
dotted line in Fig.~\ref{fig-model}b. The difference between the
observations and the CO$_2$ plus H$_2$O model is shown in
Fig.~\ref{fig-model}c which represents the CO outburst. The predicted
CO production rates during this time are shown in Table~\ref{fluxes}.

Matching water production estimates from 2013 March 5 and May 4
(\citet{schleicher13a, schleicher13b}; see Table~\ref{fluxes}),
required that the effective water-ice sublimating area was
$\sim$6$\times$ that of the nucleus surface in March dropping to
$\sim$80\% of the surface in May as the heliocentric distance
decreased.  The model run-out was also consistent with all of the
other H$_2$O production rate upper limits shown in Table~\ref{fluxes}.
As was seen for 103P/Hartley 2 \citep{ahearn11}, we propose that
this outburst ejected large water ice grains into the coma.  It has
been shown experimentally that sublimation from deeper layers can
result in the ejection of large slow-moving grains \citep{laufer05}.
A long-lived population of large ($\sim$100$\mu$m) water-ice grains
in the near nucleus environment could explain this water production
behavior.

Dirty (low albedo) ice grains of this size could survive for months
at the low temperatures outside $r$=8~AU, however inside 6~AU they
would not survive very long \citep{hanner81}.  On the other hand,
for moderately high albedo grains ($p_v$$>$0.5) lifetimes of months
to days are possible from $r$=6 to 3.5~AU (2012 September through
2013 May).  As noted in $\S$~\ref{fpmodel}, grains this large would
not contribute optically to the tail structure as they would remain
close to the nucleus in projection.  Once in the coma, the water
ice grains slowly sublimate, releasing dust into the coma and
increasing its cross section.  With high albedos and long grain
lifetimes, the duration of the brightening by this mechanism could
be many months (e.g. 2012 January-2013 January).  The fading was
not caused by loss of large grains from the aperture, grains that
would survive for the duration are so large that they stay within
the 5$''$ radius aperture (equivalent to $\sim$15,000 km) in 2013
January.  At a typical velocity of 1~m s$^{-1}$ the crossing time
for the aperture is $\sim$months.  The effective brightness decline
(compared to an expected increase from sublimation as $r$ decreased)
was thus a loss of cross section within the aperture.  With the
limited data we have, we cannot say if the outburst was short-lived,
or if there was continued activity, although possible detections
of CO after 2013 January suggest that there was outgassing for a
period of time.

\section{DISCUSSION\label{discussion}}

In formulating the concept of the Oort cloud \citep{oort51}, Oort
suggested that the dearth of returning long-period comets at large
semi-major axis was due to chemical alteration of their surface layers by
cosmic rays, and that this `volatile frosting' was lost on the first
passage through the inner solar system.  One interpretation of comet
ISON's heliocentric light curve could be its activity through 2013
January was dominated by the loss of this highly volatile layer,
and the activity since then has been decreasing.  The implication
of this interpretation is that the comet will not brighten as dramatically
as hoped near perihelion, and that the apparent brightness coming out of
solar conjunction would have remained flat or even decreased. This is
similar to the behavior observed for Comet C/1973 E1 (Kohoutek).

In the second scenario for activity where the comet is largely
driven by CO$_2$ outgassing, our models predicted that the apparent
brightness within a 5$''$ radius aperture should be $\sim$R=14-14.5
when the comet came out of solar conjunction in late August/early
September matching closely what has occured. While it is unwise to make
predictions about the brightness at perihelion when the comet is still
far from the sun, especially when it will pass so close to the sun,
the run out of these sublimation models show that the comet can
still be quite bright at perihelion.

\begin{acknowledgements}

Gemini is operated by AURA under a cooperative agreement with the
NSF on behalf of the Gemini partnership.  The James Clerk Maxell
Telescope is operated by the Joint Astronomy Centre on behalf of
the Science and Technology Facilities Council of the United Kingdom,
and the National Research Council of Canada.  Data were acquired
using the PS1 System operated by the PS1 Science Consortium (PS1SC)
and its member institutions.  The Pan-STARRS1 Surveys (PS1) have
been made possible by contributions from PS1SC member Institutions
and NASA through Grant NNX08AR22G, the NSF under Grant No. AST-123886,
the Univ. of MD, and Eotvos Lorand Univ..  B.Y., T.R. and K.J.M.
acknowledge support through the NASA Astrobiology Institute under
Cooperative Agreement NNA08DA77A.  H.H.H. is supported by NASA
through Hubble Fellowship grant HF-51274.01 awarded by the Space
Telescope Science Institute, which is operated by the Association
of Universities for Research in Astronomy for NASA, under contract
NAS 5-26555.  We also thank the following CARA observers for their
contributions: Giovanni Sostero, Ernesto Guido, Nick Howes, Martino
Nicolini, Herman Mikuz, Daniele Carosati, Jean Francois Soulier,
Man-To Hui, Gianni Galli, Walter Borghini, Paolo Bacci and Diego
Tirelli.

\end{acknowledgements}

\begin{deluxetable}{p{1.8cm}lcrcccrcccc}
\scriptsize
\tablewidth{0pt}
\tablecaption{Observations\label{obslog}}
\tablecolumns{12}
\tablehead{
\colhead{UT Date}
 & \colhead{Telescope}
 & \colhead{N\tablenotemark{a}}
 & \colhead{t\tablenotemark{b}}
 & \colhead{Filter}
 & \colhead{$r$\tablenotemark{c}}
 & \colhead{$\Delta$\tablenotemark{d}}
 & \colhead{$\alpha$\tablenotemark{e}}
 & \colhead{TA\tablenotemark{f}}
 & \colhead{PA$_{-\odot}$\tablenotemark{g}}
 & \colhead{PA$_{-v}$\tablenotemark{h}}
 & \colhead{$m_R$\tablenotemark{i}}
}
\startdata
2011-09-30 & PS1         &  4 &  180 & $w_{P1}$ & 9.392 & 9.679 &5.77&-175.83 & 284.9 & 299.8 & 20.91$\pm$0.12 \\
2011-11-10 & PS1         &  2 &   90 & $i_{P1}$ & 9.064 & 8.669 &5.87& -175.75 & 278.4 & 294.5 & 20.64$\pm$0.11 \\
2011-11-26 & PS1         &  2 &   86 & $g_{P1}$ & 8.934 & 8.302 &5.05& -175.72 & 274.2 & 293.4 & 20.42$\pm$0.10 \\
2011-12-09 & PS1         &  2 &   30 & $z_{P1}$ & 8.829 & 8.043 &4.04& -175.69 & 268.9 &  293.2 & 19.92$\pm$0.12 \\
2012-01-05 & PS1         &  1 &   43 & $g_{P1}$ & 8.606 & 7.643 &1.41& -175.63 & 226.3 & 302.9 & 19.55$\pm$0.09 \\
2012-01-28 & PS1         &  2 &   90 & $w_{P1}$ & 8.416 & 7.491 &2.44& -175.58 & 125.4 &  61.9 & 19.67$\pm$0.02 \\
2012-10-11 & HCT2m       &  5 & 1500 & R     & 6.089 & 6.212 &  9.26 & -174.80 & 284.9 & 297.2 & 17.49$\pm$0.01 \\
2012-10-14 & Calar Alto 1.2m &  7 & 1800 & R & 6.067 & 6.155 &  9.32 & -174.79 & 284.7 & 297.0 & 17.49$\pm$0.02 \\
2012-11-08 & UH2.2m      &  2 &  600 & R     & 5.814 & 5.474 &  9.46 & -174.68 & 281.1 & 294.0 & 17.12$\pm$0.01 \\
2012-11-22 & Lowell 1.8m &  3 & 1800 & VR    & 5.672 & 5.114 &  8.68 & -174.70 & 278.0 & 292.5 & 16.87$\pm$0.01 \\
2012-12-20 & Lowell 1.8m &  3 & 1800 & R     & 5.383 & 4.502 &  5.10 & -174.47 & 264.3 & 290.3 & 16.32$\pm$0.01 \\
2012-12-23 & PS1         &  4 &  180 & $w_{P1}$ & 5.350 & 4.446 &4.55& -174.46 & 261.1&  290.4& 16.35$\pm$0.03 \\
2013-01-03 & PS1         &  1 &   40 & $r_{P1}$ & 5.234 & 4.275 &2.62& -174.40 & 238.3 &  292.8 & 16.02$\pm$0.04 \\
2013-01-06 & Calar Alto 1.2m &  3 & 900 & R  & 5.199 & 4.231 &  2.15 & -174.37 & 223.3 & 295.5 & 16.08$\pm$0.03 \\
2013-01-13 & Lowell 1.8m &  2 & 1200 & R     & 5.129 & 4.157 &  1.87 & -174.34 & 177.9 & 313.7 & 16.81$\pm$0.01 \\
2013-02-04 & Gemini N 8m &  2 &  150 & $r$   & 4.891 & 4.019 &  5.95 & -174.20 & 113.2 &  85.6 & 15.88$\pm$0.10 \\
2013-03-04 & Gemini N 8m &  2 &   90 & $r$   & 4.578 & 4.050 & 11.19 & -174.01 &  99.1 &  84.4 & 15.90$\pm$0.10 \\
2013-04-03 & Gemini N 8m &  4 &  197 & $r$   & 4.231 & 4.206 & 13.61 & -173.77 &  93.0 &  80.3 & 15.87$\pm$0.10 \\
2013-05-01 & Calar Alto 1.2m & 4 & 1200 & R  & 3.886 & 4.326 & 12.70 & -173.50 &  90.2 &  75.6 & 15.92$\pm$0.02 \\
2013-05-04 & Gemini N 8m &  3 &  135 & $gri$ & 3.857 & 4.331 & 12.50 & -173.48 & 90.0 &  75.1 & 15.61$\pm$0.10 \\
2013-05-17 & UH2.2m      &  4 & 1200 & R     & 3.693 & 4.341 & 11.13 & -173.42 & 89.6 &  73.6 & 15.85$\pm$0.01 \\
2013-05-30 & Gemini N 8m &  2 &   90 & $r$   & 3.528 & 4.318 &  9.34 & -173.18 & 87.6 & 66.2 & 15.65$\pm$0.01 \\
2013-08-12 & G95\tablenotemark{j} 11-in &    &      & R     & 2.487 & 3.418 &  7.88 & -171.89 & 294.2 & 311.3 & 14.73$\pm$0.10 \\
2013-08-16 & G95         &    &      & R     & 2.425 & 3.332 &  9.07 & -171.79 & 293.1 & 309.1 & 14.71$\pm$0.10 \\
\hline
UT Date & 
Telescope &
Mol. & 
Line & 
$f\tablenotemark{k}$ & 
t$_{int}\tablenotemark{l}$ & 
B$_{eff}$ \tablenotemark{m}& 
$r$ & 
$\Delta$ & 
rms\tablenotemark{n} & 
T\tablenotemark{p} & 
Q$_{CO}$\tablenotemark{q} \\
\hline
2013-03-10,15    & JCMT & CO & J(3-2) & 345.8 &  7500 & 0.52 & 4.47 & 4.09 & 6.1& 18 & $<$3.9 \\
2013-03-30       & JCMT & CO & J(3-2) & 345.8 &  6468 & 0.50 & 4.26 & 4.19 & 5.5& 19 & $<$3.5 \\
2013-04-01       & JCMT & CO & J(3-2) & 345.8 &  6468 & 0.50 & 4.26 & 4.19 & 5.5& 19 & $<$3.5 \\
2013-04-27,28    & JCMT & CO & J(2-1) & 230.5 &  9156 & 0.50 & 3.93 & 4.32 & 4.6& 21 & $<$4.5 \\
2013-06-14,15    & JCMT & CO & J(2-1) & 230.5 & 15042 & 0.50 & 3.32 & 4.23 & 3.2& 26 & $<$4.1 \\
\enddata
\tablenotetext{a}{Number of exposures.}
\tablenotetext{b}{Total integration time, s.}
\tablenotetext{c}{Heliocentric distance, AU.}
\tablenotetext{d}{Geocentric distance, AU.}
\tablenotetext{e}{Solar phase angle, degrees.}
\tablenotetext{f}{True anomaly, degrees.}
\tablenotetext{g}{Position angle of the antisolar vector, degrees East of North.}
\tablenotetext{h}{Position angle of the negative velocity vector, degrees East of North.}
\tablenotetext{i}{Mean apparent R-band magnitude.}
\tablenotetext{j}{Hereford Arizona Observatory}
\tablenotetext{k}{Rest Frequency, in GHz.}
\tablenotetext{l}{Total integration time, s.}
\tablenotetext{m}{Main Beam Efficiency.}
\tablenotetext{n}{1-$\sigma$ rms noise in unit of the main beam brightness temperature, in mK. } 
\tablenotetext{p}{Kinetic temperature computed based on the formula from \cite{biver97}, in K.}
\tablenotetext{q}{3-$\sigma$ production rates upper limits for CO, in 10$^{27}$ molec s$^{-1}$.}
\end{deluxetable}

\begin{deluxetable}{p{1.8cm}lcrcccrcccc}
\scriptsize
\tablewidth{0pt}
\tablecaption{CARA and VYSOS Observations\label{obslog2}}
\tablecolumns{12}
\tablehead{
\colhead{UT Date}
 & \colhead{Tel.\tablenotemark{a}}
 & \colhead{N}
 & \colhead{t}
 & \colhead{Filter}
 & \colhead{$r$}
 & \colhead{$\Delta$}
 & \colhead{$\alpha$}
 & \colhead{TA}
 & \colhead{PA$_{-\odot}$}
 & \colhead{PA$_{-v}$}
 & \colhead{$m_R$}
}
\startdata
2012-09-25 & CARA-Net &  &  & R & 6.250 & 6.636 & 8.240 & -174.868 & 286.8 & 299.7 & 17.69$\pm$0.21 \\
2012-10-04 & CARA-Net &  &  & R & 6.161 & 6.405 & 8.850 & -174.831 & 285.8 & 298.3 & 17.05$\pm$0.10 \\
2012-10-10 & CARA-Net &  &  & R & 6.103 & 6.251 & 9.167 & -174.807 & 285.1 & 297.4 & 17.10$\pm$0.20 \\
2012-10-22 & CARA-Net &  &  & R & 5.985 & 5.931 & 9.566 & -174.756 & 283.6 & 295.9 & 17.01$\pm$0.17 \\
2012-10-23 & CARA-Net &  &  & R & 5.978 & 5.912 & 9.579 & -174.753 & 283.5 & 295.8 & 17.07$\pm$0.20 \\
2013-04-02 & VYSOS & 3 & 300 & r & 4.242 & 4.201 & 13.589 & -173.78 & 93.2 & 80.4 & 16.10$\pm$0.06 \\
2013-04-03 & VYSOS & 6 & 600 & r & 4.232 & 4.205 & 13.608 & -173.77 & 93.0 & 80.3 & 15.68$\pm$0.02 \\
2013-04-04 & VYSOS & 3 & 300 & r & 4.220 & 4.211 & 13.627 & -173.76 & 92.9 & 80.1 & 15.89$\pm$0.04 \\
2013-04-05 & VYSOS & 3 & 300 & r & 4.208 & 4.216 & 13.642 & -173.75 & 92.8 & 80.0 & 15.90$\pm$0.04 \\
2013-04-06 & VYSOS & 3 & 300 & r & 4.196 & 4.222 & 13.653 & -173.75 & 92.7 & 79.9 & 15.89$\pm$0.05 \\
\enddata
\tablenotetext{a}{See notes for Table~\ref{obslog}}
\tablenotetext{b}{Table 2 is published in its entirety in the
electronic edition.  A portion is shown here for guidance regarding
its form and content.}
\end{deluxetable}


\begin{deluxetable}{p{2.5cm}ccccccllc}
\scriptsize
\tablewidth{0pt}
\tablecaption{2013 Volatile Flux Constraints\label{fluxes}}
\tablecolumns{10}
\tablehead{
\colhead{UT Date 2013}
 & \colhead{JD\tablenotemark{a}}
 & \colhead{$r$\tablenotemark{b}}
 & \colhead{TA\tablenotemark{c}}
 & \colhead{$Q_{CO}$\tablenotemark{d}}
 & \colhead{$Q_{CO2}$\tablenotemark{d}}
 & \colhead{$Q_{H2O}$\tablenotemark{d}}
 & \colhead{Facility}
 & \colhead{Ref\tablenotemark{e}}
 & \colhead{Model $Q_{CO}$\tablenotemark{f}}
}
\startdata
January 30   & 56323& 4.94& -174.23&           & & $<$1E28 &Swift   & 1& \\
March 5      & 56356& 4.57& -174.00&           & & 3E26    &Lowell  & 2& \\
March 11     & 56363& 4.50& -173.96& 	       & & $<$1E28 &Swift   & 1& \\
March 10-15  & 56365& 4.52& -173.97& $<$3.9E27 & &         &JCMT    & 3&1.9E27\\
March 30-April 1& 56382& 4.28& -173.80& $<$3.5E27 & &         &JCMT    & 3&1.7E27\\
April 24     & 56407& 3.98& -173.58&           & & $<$1E28 &Swift   & 1& \\
April 27-28  & 56410& 3.95& -173.55& $<$4.5E27 & &         &JCMT    & 3&1.5E27\\
May 4        & 56417& 3.86& -173.48& 	       & & 6E26    &Lowell  & 5& \\
May 9        & 56422& 3.79& -173.42& 	       & & $<$1E28 &Swift   & 1& \\
June 13	     & 56457& 3.34& -173.00& 2.1E27    & 1.9E26 &  &Spitzer & 6&7.0E26\\
June 14-15   & 56458& 3.33& -172.98& $<$4.1E27 & &         &JCMT    & 3&7.0E26\\
\enddata
\tablenotetext{a}{Julian date-2400000.}
\tablenotetext{b}{Heliocentric distance, AU.}
\tablenotetext{c}{True anomaly, degrees.}
\tablenotetext{d}{Production rate, molec s$^{-1}$.}
\tablenotetext{e}{References: 1-\citep{bodewits13}, 
2-\citep{schleicher13a}, 3-this paper, 
4-A'Hearn private comm, 5-\citep{schleicher13b}, 6-\citep{lisse13}.}
\tablenotetext{f}{CO production rate prediction from model CO outburst.}
\end{deluxetable}


\clearpage

\end{document}